\documentclass[prl,twocolumn,english,superscriptaddress,amsmath,amssymb,floatfix]{revtex4-2}

\usepackage{float}
\usepackage{graphicx}
\usepackage{dcolumn}
\usepackage{bm}

\usepackage{color}
\usepackage[colorlinks,bookmarks=false,citecolor=darkblue,linkcolor=darkblue,urlcolor=blue]{hyperref}

\definecolor{darkred}{rgb}{0.7,0.0,0.0}

\definecolor{darkblue}{rgb}{0,0.02,0.45}

\definecolor{darkgreen}{rgb}{0.02,0.45,0.0}

\definecolor{violet}{rgb}{0.8,0.2,0.6}

\graphicspath{{graphs/}}

\begin{document}

\title{Thermodynamic signatures of a nearby quantum critical endpoint\\ in UTe$_2$ under intense magnetic field}

\author{M. Gen}
\thanks{These authors contributed equally to this work.}
\affiliation{Institute for Solid State Physics, University of Tokyo, Kashiwa 277-8581, Japan}
\author{K. Somesh}
\thanks{These authors contributed equally to this work.}
\affiliation{Laboratoire National des Champs Magn\'{e}tiques Intenses - EMFL, CNRS, Universit\'{e} Toulouse, Universit\'{e} Grenoble Alpes, INSA-T, Toulouse, France}
\author{A. Miyake}
\affiliation{Institute for Materials Research, Tohoku University, Ibaraki 311-1313, Japan}
\author{D. Li}
\affiliation{Institute for Materials Research, Tohoku University, Ibaraki 311-1313, Japan}
\author{Z. Yang}
\affiliation{Institute for Solid State Physics, University of Tokyo, Kashiwa 277-8581, Japan}
\author{A. Miyata}
\affiliation{Institute for Solid State Physics, University of Tokyo, Kashiwa 277-8581, Japan}
\author{Y. Kohama}
\affiliation{Institute for Solid State Physics, University of Tokyo, Kashiwa 277-8581, Japan}
\author{M. Kondo}
\affiliation{Institute for Solid State Physics, University of Tokyo, Kashiwa 277-8581, Japan}
\author{M. Tokunaga}
\affiliation{Institute for Solid State Physics, University of Tokyo, Kashiwa 277-8581, Japan}
\author{D. Aoki}
\affiliation{Institute for Materials Research, Tohoku University, Ibaraki 311-1313, Japan}
\author{W. Knafo}
\affiliation{Laboratoire National des Champs Magn\'{e}tiques Intenses - EMFL, CNRS, Universit\'{e} Toulouse, Universit\'{e} Grenoble Alpes, INSA-T, Toulouse, France}
\affiliation{Institute for Materials Research, Tohoku University, Ibaraki 311-1313, Japan}

\date{\today}

\begin{abstract}

Several unconventional superconducting phases have been discovered in the vicinity of metamagnetism in the heavy-fermion paramagnet UTe$_2$ under magnetic field. To describe the mechanism of magnetic-field-induced superconductivity, a challenge is to characterize the quantum critical magnetic properties at the metamagnetic transition. In this work, we investigate the thermodynamic properties of UTe$_2$ under magnetic fields up to 60~T applied in the crystallographic $\mathbf{b}-\mathbf{c}$ plane. Magnetization and magnetostriction experiments have been performed under magnetic fields tilted by an angle $\theta\lesssim42^\circ$ from $\mathbf{b}$ toward $\mathbf{c}$. They support that a quantum critical endpoint at the field angle $\theta_{QCEP}\simeq(70\pm10)^\circ$ and magnetic field $\mu_0H_{QCEP}\simeq(75\pm10)$~T terminates the metamagnetic transition line. A pseudo-gap in the magnetic susceptibility is related with the metamagnetic transition observed in magnetic fields for $\theta<\theta_{QCEP}$. The possible relation between the quantum critical endpoint and the development of magnetic-field-induced superconductivity is discussed.

\end{abstract}

\maketitle

\begin{figure}[b]
\centering
\includegraphics[width=0.5\textwidth]{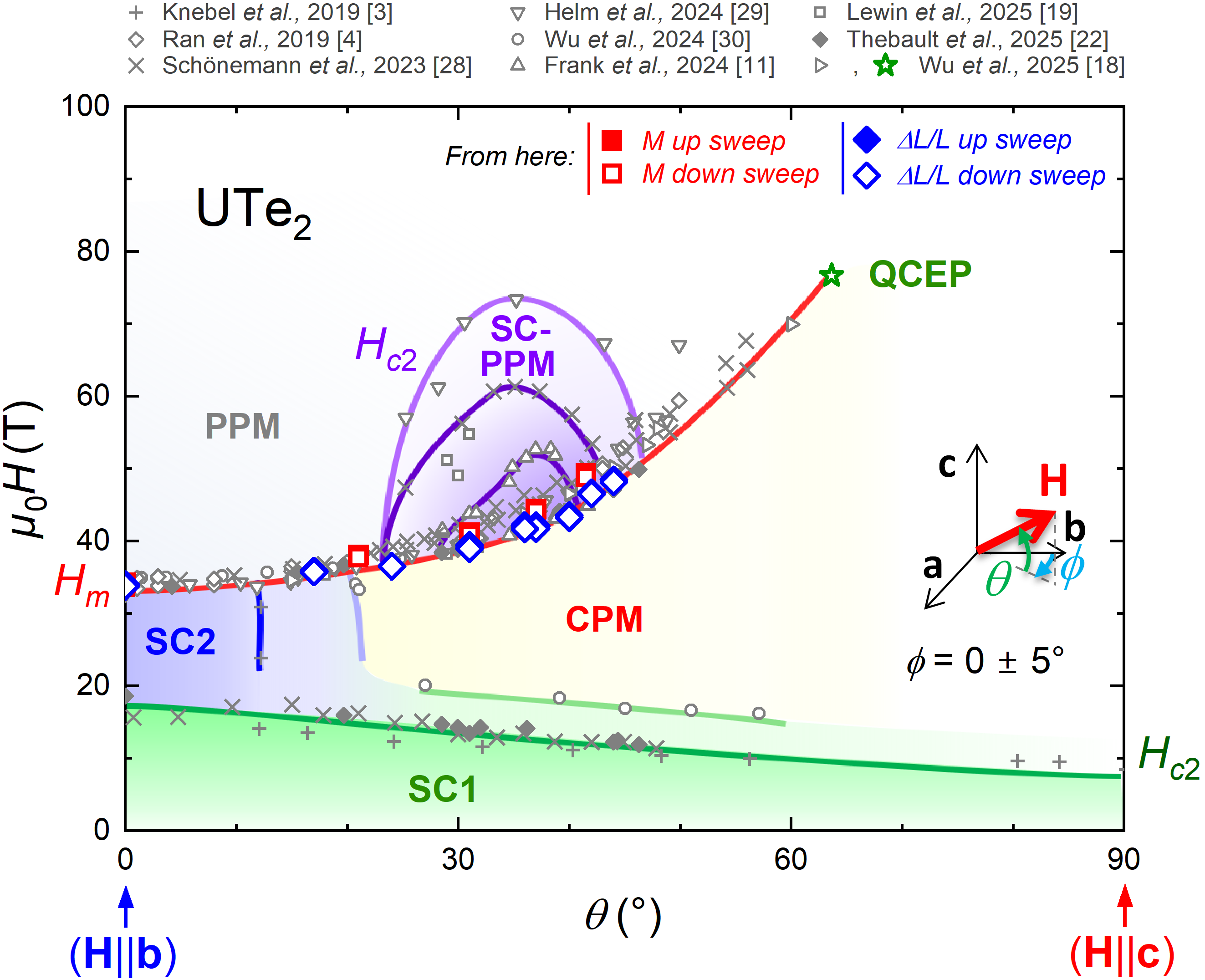}
\caption{\label{Fig1} Field angle $\theta$ versus magnetic field strength $H$ phase diagram of UTe$_2$ under magnetic fields tilted by $\theta$ from $\mathbf{b}$ to $\mathbf{c}$, constructed using data from here and \cite{Knebel2019,Ran2019b,Schonemann2024,Helm2024,Frank2024,Wu2024,Wu2025,Lewin2025,Thebault2026}. SC1, SC2, and SC-PPM denote the superconducting phases, CPM and PPM denote the correlated paramagnetic and polarized paramagnetic regimes, respectively, and QCEP denotes a quantum critical endpoint.}
\end{figure}

An unprecedented number of unconventional superconducting phases, often suspected to be triplet, have been discovered in UTe$_2$ in the vicinity of quantum magnetic phase transitions tuned by pressure and magnetic field \cite{Ran2019a,Aoki2019a,Knebel2019,Ran2019b,Braithwaite2019,Knafo2021a,Valiska2021,Aoki2021,Kinjo2023,Honda2023,Frank2024}. In particular, two superconducting phases SC2 and SC-PPM are induced by a magnetic field $H$ near a metamagnetic transition to a polarized paramagnetic (PPM) regime (see Figure \ref{Fig1}). For $\mathbf{H}\parallel\mathbf{b}$, the metamagnetic transition at $\mu_0H_m=34$~T has a first-order character driving step-like variations in the magnetization $M$ \cite{Miyake2019,Ran2019b,Miyake2021}, sample lengths $\Delta L$ measured along the directions $\mathbf{a}$, $\mathbf{b}$, and $\mathbf{c}$ \cite{Miyake2022}, and electrical resistivity \cite{Knafo2019,Thebault2022}. SC2 is observed only in the correlated paramagnetic (CPM) regime under magnetic fields $H<H_m$ applied along the direction $\mathbf{b}$, and it rapidly disappears in fields tilted by a few degrees from $\mathbf{b}$ \cite{Knebel2019,Lewin2024}. A tilt of the magnetic field away from $\mathbf{b}$ also leads to an increase of the metamagnetic field $H_m$ \cite{Ran2019b}. In magnetic fields $\mathbf{H}$ tilted by angles $\theta\simeq30-40^\circ$ from $\mathbf{b}$ to $\mathbf{c}$, the superconducting phase SC-PPM is observed in the high-field PPM regime reached for $\mu_0H>\mu_0H_m\simeq40-45$~T \cite{Ran2019b,Knafo2021a}. SC-PPM is stabilized within an unusual three-dimensional domain, a distorted halo bounded at low fields by the metamagnetic surface \cite{Wu2025,Lewin2025}. Quantum critical magnetic fluctuations induced at $H_m$, whose signatures have been observed by microscopic and macroscopic probes \cite{Knafo2019,Knafo2021a,Rosuel2023a,Thebault2022,Tokunaga2023,Thebault2026}, are suspected to drive the mechanisms of SC2 and SC-PPM (see also \cite{Duan2020,Knafo2021b,Duan2021,Raymond2021}). In particular, asymmetric field variations of Fermi-liquid coefficients through $H_m$ indicate stronger magnetic fluctuations for $H\lesssim H_m$ with $\mathbf{H}\parallel\mathbf{b}$ and for $H\gtrsim H_m$ with $\mathbf{H}$ tilted by $\theta\simeq30-40^\circ$, which coincides with the domains of stability of SC2 and SC-PPM, respectively \cite{Rosuel2023a,Thebault2022,Thebault2026}. Recent proximity-detector-oscillator (PDO) measurements indicated that the first-order metamagnetic transition, at which the frequency shift $\Delta f_{PDO}$ shows a step-like variation, terminates into a quantum critical end point (QCEP) at $\theta_{QCEP}\simeq64^\circ$ and $\mu_0H_{QCEP}\simeq77$~T when the magnetic field is tilted from $\mathbf{b}$ to $\mathbf{c}$ \cite{Wu2025} (see Figure \ref{Fig1}). Non-absolute magnetization measurements supported that the amplitude of the first-order peak at $H_m$ in $\partial M/\partial H$ decreases with increasing $\theta$ before disappearing for $\theta>\theta_{QCEP}\simeq60^\circ$ \cite{Czajka2025}. However, the amplitude of the jump at $H_m$ in the magnetic torque $\tau$ of UTe$_2$ was shown to increase, either moderately \cite{Schonemann2024} or strongly \cite{Helm2024}, with increasing field angles $\theta$ up to $\simeq55^\circ$. These different works indicate that the presence of a QCEP in UTe$_2$ under magnetic fields tilted away from the direction $\mathbf{b}$ has not been demonstrated unambiguously so far. Knowing that a first-order transition is defined at a step-like variation in the first-derivatives of the free energy, as the magnetization, the length, and the entropy, a study using absolute thermodynamic probes is mandatory to definitively prove that a QCEP is established at the termination of the first-order-transition metamagnetic line of UTe$_2$, but also to try understanding its possible relation with the establishment of magnetic-field-induced superconductivity near the metamagnetic transition.

In this work, we have studied the metamagnetic transition of UTe$_2$ by combining absolute magnetization and magnetostriction measurements under pulsed magnetic fields tilted from $\mathbf{b}$ to $\mathbf{c}$ and up to 60~T. Complementary low-field magnetic-susceptibility measurements have also been performed. Decreases of the jumps $\Delta M$ and $\Delta(\Delta L_b/L_b)$ at the metamagnetic transition in the magnetization and relative length measured along the direction $\mathbf{b}$, respectively, are observed under magnetic fields tilted from $\mathbf{b}$ to $\mathbf{c}$. They are related with the progressive closing of a pseudo-gap in the low-temperature and low-field magnetic susceptibility $\chi$. These effects are the signature of a nearby QCEP terminating the metamagnetic-transition line. The possible relationship between this QCEP and unconventional superconductivity induced near $H_m$ under tilted magnetic fields is discussed.

The samples investigated here have been grown by the chemical-vapor-transport technique. Samples of dimensions (mass) 1.5*0.5*1.5 mm$^3$ (14.1~mg), 1.5*2*0.5 mm$^3$ (18.8~mg), and 2*2.5*0.4 mm$^3$ (24.6~mg) have been studied by magnetization, magnetostriction and magnetic susceptibility, respectively, under pulsed or steady magnetic fields applied in the plane ($\mathbf{b}$,$\mathbf{c}$). Pulsed magnetic fields have been generated at the International MegaGauss Science Laboratory in Kashiwa. Magnetization in pulsed magnetic fields up to 60~T was measured using the compensated-coil technique. Magnetostriction from sample length $\mathbf{L_b}$ along the direction $\mathbf{b}$ was measured under pulsed magnetic fields up to 60~T using the fiber-Bragg-grating strain-gage technique. Spectra were obtained using a spectrometer with InGaAs camera (NIRvana HS, Teledyne Princeton Instruments) with an acquisition rate of $\simeq5.2$kHz, which allows a more reliable determination of the length change than the optical-filter-based method (see for instance \cite{Ikeda2018}). Magnetostriction data have been collected for the different field directions without detaching the optical fiber from the sample. Magnetic susceptibility in a steady magnetic field of 1~T was measured using a commercial MPMS SQUID magnetometer from Quantum Design.

Figure \ref{Fig1} presents the field angle $\theta$ versus magnetic-field strength $H$ phase diagram of UTe$_2$ under magnetic fields tilted by $\theta$ from $\mathbf{b}$ to $\mathbf{c}$, constructed using data from here and \cite{Knebel2019,Ran2019b,Schonemann2024,Helm2024,Frank2024,Wu2024,Wu2025,Lewin2025,Thebault2026}. The angular dependence of the metamagnetic field $H_m$ is weakly sensitive to the sample quality \cite{Frank2024}, and small differences in the values of $H_m$ probably result from small additional - and unintended - tilts of the magnetic field toward the direction $\mathbf{a}$ \cite{Ran2019b,Wu2024,Lewin2024,Wu2025,Lewin2025,Czajka2025}. $H_m$ delimitates the low-field CPM regime with strong electronic correlations and weakly-polarized magnetic moments from a high-field PPM regime with weak electronic correlations and strongly-polarized magnetic moments. The boundaries of the superconducting phases, SC1 in low field, SC2 and SC-PPM in high field, are strongly affected by the sample quality and larger regions of stability of the superconducting phases are reported in the phase diagrams of higher-quality samples \cite{Knebel2019,Wu2024,Helm2024,Schonemann2024,Lewin2025}.

\begin{figure}[!t]
\centering
\includegraphics[width=0.47\textwidth]{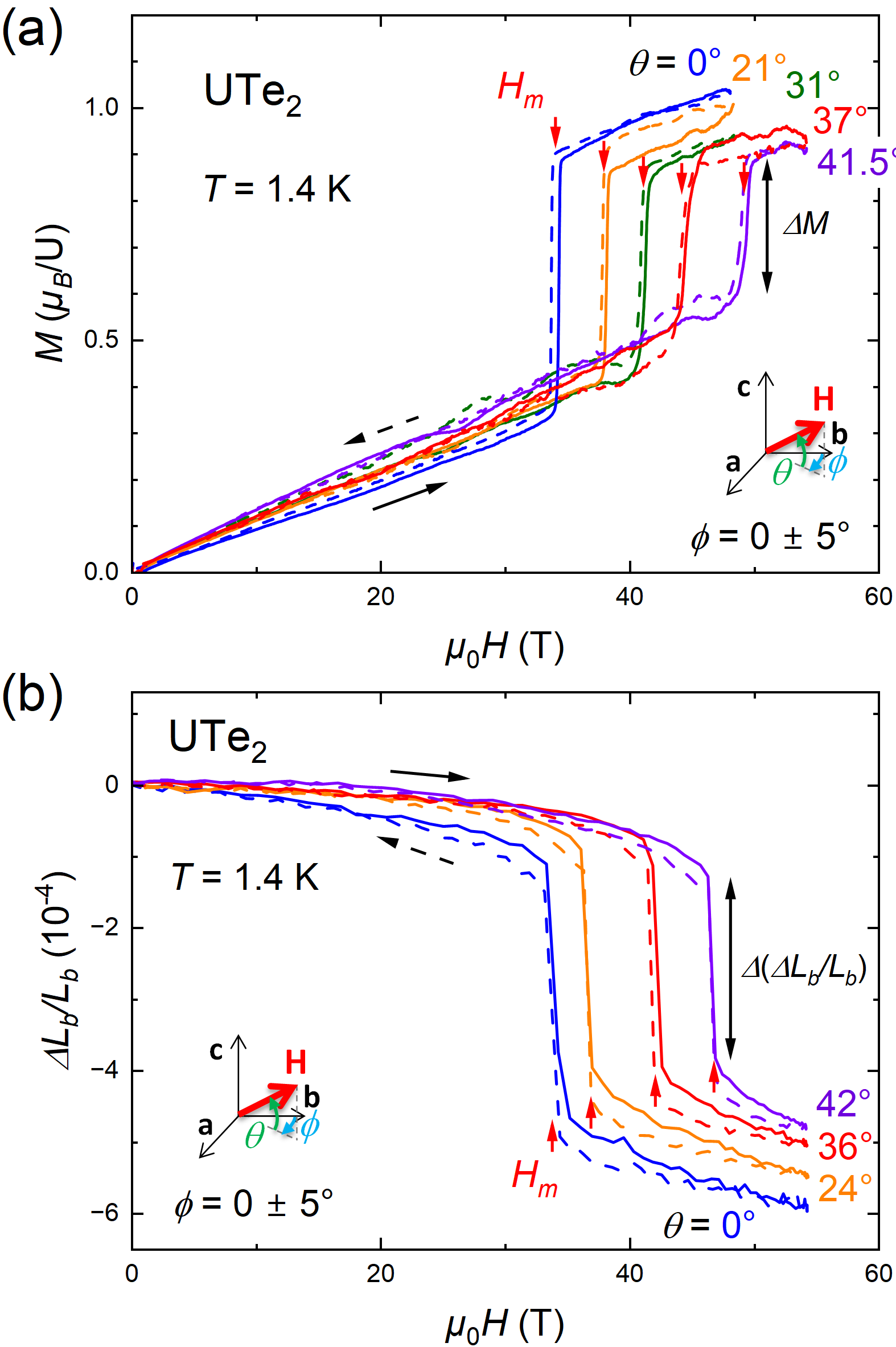}
\caption{\label{Fig2} Magnetic-field variation of the (a) magnetization $M$ and (b) relative length $\Delta L_b/L_b$, where $L_b$ is the sample length along the direction $\mathbf{b}$ and $\Delta L_b=L_b-L_b(H=0)$, measured at field angles $\theta$ from 0 to $42^\circ$ and temperature $T=1.4$~K, for up (full lines) and down (dashed lines) sweeps of pulsed magnetic fields.}
\end{figure}

\begin{figure}[!t]
\centering
\includegraphics[width=0.47\textwidth]{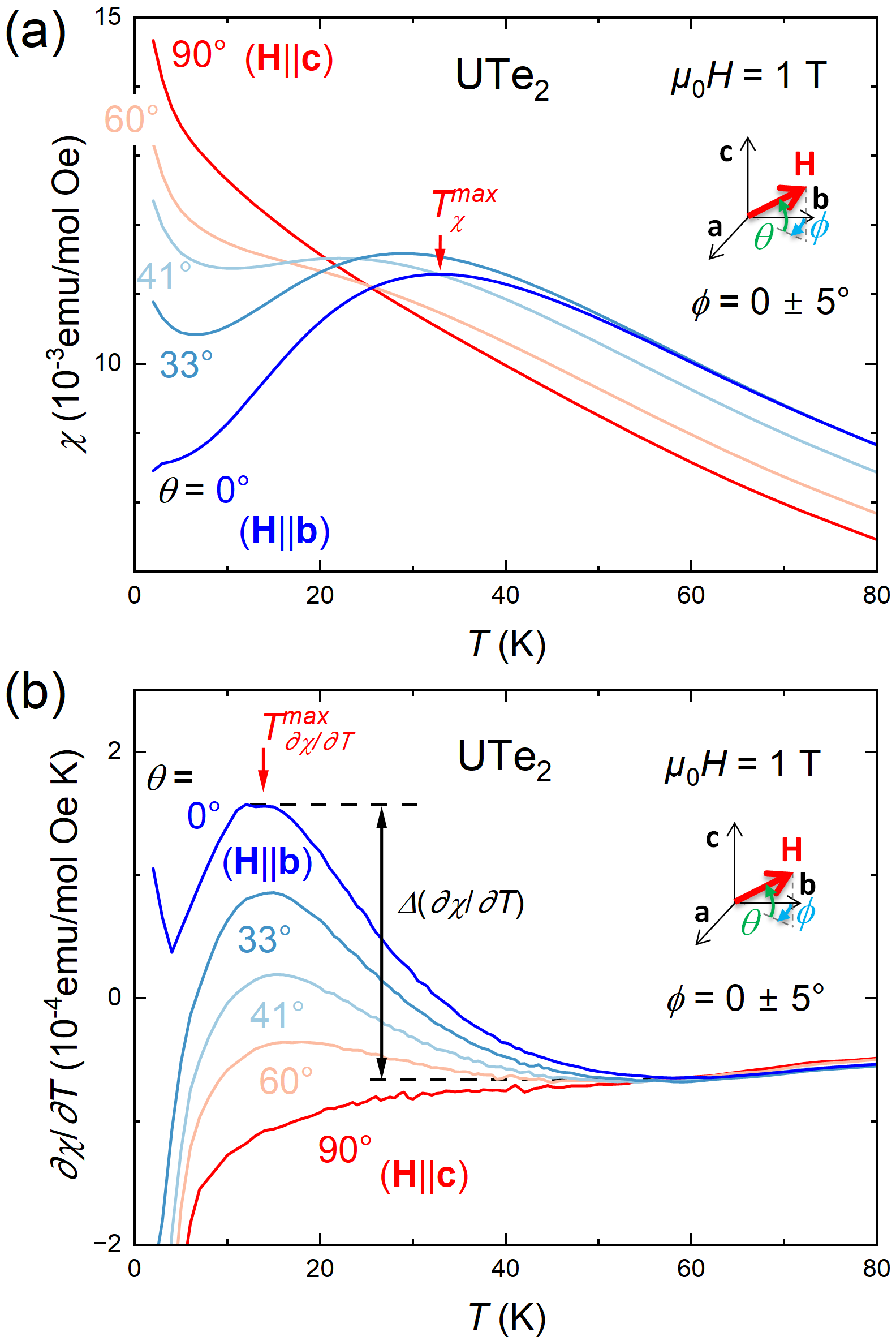}
\caption{\label{Fig3} Temperature dependence of (a) the magnetic susceptibility $\chi=M/H$ and (b) its temperature derivative $\partial\chi/\partial T$ measured in magnetic fields of $1$~T tilted by angles $0\leq\theta\leq90^\circ$ from $\textbf{b}$ to $\textbf{c}$.}
\end{figure}

Figures \ref{Fig2}(a-b) show the magnetic-field variations of the magnetization $M$ and relative length $\Delta L_b/L_b$, where $\Delta L_b=L_b-L_b(H=0)$, measured at field angles $\theta$ from 0 to $42^\circ$ and temperature $T=1.4$~K, for up- and down-sweeps of pulsed magnetic fields. The increase of $\mu_0H_m$ versus $\theta$, up to $\simeq50$~T for $\theta\simeq42^\circ$, observed in both sets of data is similar to that from previously published works \cite{Ran2019b,Schonemann2024,Helm2024,Frank2024,Wu2024,Wu2025,Lewin2025,Thebault2026} (see Figure \ref{Fig1}). Slightly larger values of $H_m$ in our magnetization data at field angles $\theta>20^\circ$ probably result from small additional tilts of the magnetic field toward the direction $\mathbf{a}$. Figure \ref{Fig2} shows that the absolute values of the amplitudes of the steps $\Delta M$ in the magnetization and $\Delta(\Delta L_b/L_b)$ in the relative length at the first-order transition at $H_m$ decrease with increasing field angle $\theta$. $\Delta M$ at $H_m$ varies from $0.54\pm0.01$~$\mu_B/$U for $\theta=0$ to $0.30\pm0.02$~$\mu_B/$U for $\theta=41.5^\circ$ [see Figure \ref{Fig2}(a)], while $\Delta(\Delta L_b/L_b)$ at $H_m$ varies from $(-3.35\pm0.1)\times10^{-4}$ for $\theta=0^\circ$ to  $(-2.6\pm0.1)\times10^{-4}$ for $\theta=42^\circ$ [see Figure \ref{Fig2}(b)].

Figure \ref{Fig3}(a) shows the temperature dependence of the magnetic susceptibility $\chi=M/H$ measured in magnetic fields of $1$~T tilted by angles $0\leq\theta\leq90^\circ$ from $\textbf{b}$ to $\textbf{c}$. For $\mathbf{H}\parallel\mathbf{b}$ ($\theta=0^\circ$), $\chi$ presents a broad maximum at the temperature $T_\chi^{max}\simeq33$~K marking the onset of low-temperature magnetic correlations. $\chi$ decreases from $11.3~\times10^{-3}$~emu/mol~Oe at $T_\chi^{max}$ to $8.5~\times10^{-3}$~emu/mol~Oe at $T=2$~K. With increasing $\theta$, the amplitude of the anomaly at $T_\chi^{max}$ decreases, indicating a lower sensitivity of $\chi$ to the magnetic correlations, and a low-temperature upturn develops in $\chi$. For $\mathbf{H}\parallel\mathbf{c}$ ($\theta=90^\circ$), the signatures of low-temperature magnetic correlations have disappeared and $\chi$ increases monotonously when the temperature decreases. Figure \ref{Fig3}(b) presents the temperature-derivative of the magnetic susceptibility $\partial\chi/\partial T$ versus temperature. At field angles $\theta\leq60^\circ$, an enhancement of $\partial\chi/\partial T$ at $T_{\partial\chi/\partial T}^{max}$, corresponding to the inflexion point preceding the maximum of $\chi$ versus $T$, is a signature of the low-temperature magnetic correlations. $T_{\partial\chi/\partial T}^{max}\simeq13-15$~K is almost independent of $\theta$. The amplitude $\Delta(\partial\chi/\partial T)$ of the anomaly is estimated from the difference between $\partial\chi/\partial T$ at $T_{\partial\chi/\partial T}^{max}$ and $\partial\chi/\partial T$ at $T\simeq50-60$~K. $\Delta(\partial\chi/\partial T)$ decreases with increasing $\theta$ from $2.2~\times10^{-4}$~emu/mol~Oe~K for $\mathbf{H}\parallel\mathbf{b}$ ($\theta=0^\circ$) to $3.1~\times10^{-5}$~emu/mol~Oe~K for $\theta=60^\circ$, the anomaly being absent for $\mathbf{H}\parallel\mathbf{c}$ ($\theta=90^\circ$), where $\partial\chi/\partial T$ increases monotonously with $T$.

\begin{figure}[!b]
\centering
\includegraphics[width=0.5\textwidth]{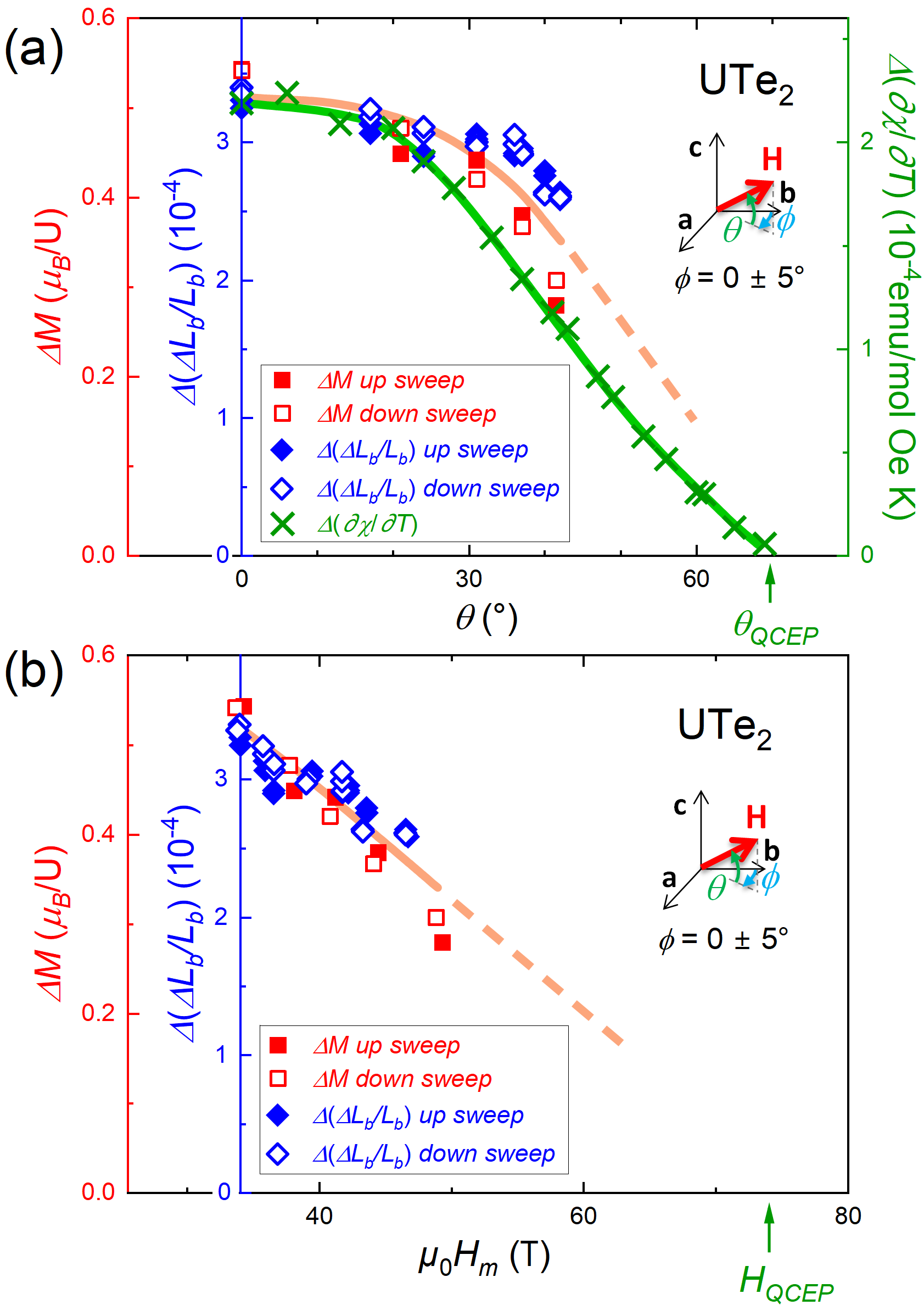}
\caption{\label{Fig4} (a) Variations with field angle $\theta$ of the jumps in the magnetization $\Delta M$ and in the relative length $\Delta(\Delta L_b/L_b)$ at $H_m$, and in the jump in the temperature derivative of the magnetic susceptibility $\Delta(\partial\chi/\partial T)$ at $T_{\partial\chi/\partial T}^{max}$. (b) Variations with metamagnetic field $H_m$ of $\Delta M$ and $\Delta(\Delta L_b/L_b)$.}
\end{figure}

Figure \ref{Fig4}(a) presents a comparison between the variations of $\Delta M$, $\Delta(\Delta L_b/L_b)$, and $\Delta(\partial\chi/\partial T)$ versus $\theta$ extracted from the data shown in Figures \ref{Fig2} and \ref{Fig3} (see also Supplemental Material \cite{SM}). These three quantities decrease continuously with increasing $\theta$: $\Delta(\partial\chi/\partial T)$ reaches zero for $\theta\simeq70^\circ$, while $\Delta M$ and $\Delta(\Delta L_b/L_b)$ have been reduced by 45 and 20~$\%$, respectively, at $\theta\simeq42^\circ$ ($H_m$ could not be investigated at larger field angles $\theta$ due to the 60-T maximum field of the magnets used here). Knowing that the metamagnetic transition is first-order, the Clapeyron relationship permits to extract $\partial H_m/\partial p_b = V_m\Delta(\Delta L_b/L_b)/\Delta M$, where $\mathbf{p}_b$ is a uniaxial pressure along the direction $\mathbf{b}$ and $V_m$ is the molar volume of UTe$_2$. For $\mathbf{H}\parallel\mathbf{b}$, from $V_m=5.38~\times10^{-5}$~m$^3$/mol, $\Delta M=0.54\mu_B/U$, and $\Delta(\Delta L_b/L_b)=-3.35~\times10^{-3}$, we extract $\partial H_m/\partial p_b=-6.0$~T/GPa, which indicates that a uniaxial pressure applied along the direction $\mathbf{b}$ leads to a decrease of $H_m$. The different field-angle-$\theta$ dependencies of $\Delta M$ and $\Delta(\Delta L_b/L_b)$ observed in Figure \ref{Fig4}(a) could in principle result from a variation of $\partial H_m/\partial p_b$ with $\theta$. However, Figure \ref{Fig4}(b) shows that the variations of $\Delta M$ and $\Delta(\Delta L_b/L_b)$ as a function of $H_m$ are quite similar, indicating that the tuning of $H_m$ may be the main effect in play. The faster variation of $\Delta M$ versus $\theta$, in comparison with that of $\Delta(\Delta L_b/L_b)$ versus $\theta$, may be - at least partly - due to a small additional tilt of the magnetic field toward the direction $\mathbf{a}$, which was suspected from the larger values of $H_m$ in our magnetization experiment (see Figures \ref{Fig1} and \ref{Fig2}). In Figure \ref{Fig4}, linear extrapolations of the variations of $\Delta M$ and $\Delta(\Delta L_b/L_b)$ with $\theta$ and $H_m$ indicate that their cancelation is expected at $\theta_{QCEP}\simeq(70\pm10)^\circ$ and $\mu_0H_{QCEP}\simeq(75\pm10)$~T. Using absolute thermodynamic probes, we therefore confirm the presence of the QCEP initially indicated from PDO \cite{Wu2025} and non-absolute magnetization \cite{Czajka2025} experiments.

In most of heavy-fermion paramagnets, the magnetic susceptibility $\chi$ presents a broad maximum at the temperature $T_\chi^{max}$ under magnetic fields of specific directions \cite{Aoki2013,Knafo2021c}. For $T<T_\chi^{max}$, a CPM regime is established and a pseudo-gap-type decrease of $\chi$ is driven by the onset of intersite magnetic fluctuations. In UTe$_2$, the maximum of $\chi$ at $T_\chi^{max}$ for $\mathbf{H}\parallel\mathbf{b}$ is related with a maximum in the electrical resistivity $\rho_{xx}$ measured with a current $\mathbf{I}\parallel\mathbf{a}$ \cite{Thebault2022,Miyake2019,Miyake2021}. We have shown here that the onset of the CPM regime leads to a broad maximum in $\partial\chi/\partial T$ at the temperature $T_{\partial\chi/\partial T}^{max}\simeq13-15$~K~$\simeq T_\chi^{max}/2$. Extrema at a similar temperature scale $T^*\simeq15$~K were reported in the electronic heat capacity $C_p$ and thermal expansion coefficients $\alpha$ \cite{Willa2021} which, as well as $\partial\chi/\partial T=H^{-1}\partial M/\partial T$, are second-degree derivatives of the free energy. The anomalies at $T_{\partial\chi/\partial T}^{max}$ and $T^*$ are presumably driven by the progressive onset of magnetic fluctuations. The CPM regime is delimited at low fields by the temperature $T_\chi^{max}$ and at high fields by the metamagnetic field $H_m$. At the temperature $T_{CEP}\simeq7$~K, a critical endpoint terminates the first-order transition line at $H_m$, which is replaced by a crossover for $T>T_{CEP}$ ending at $T_\chi^{max}\simeq35$~K in the low-field limit \cite{Knafo2019,Miyake2019,Miyake2021}. A universal law $T_\chi^{max}\sim H_m$ also indicates that these two quantities are driven by a single parameter for a large number of heavy-fermion paramagnets \cite{Aoki2013,Knafo2021c}. In the PPM regime established for $H>H_m$, the low-temperature intersite fluctuations become quenched (see for instance \cite{Rossat1988}) and $\chi$ increases monotonously when the temperature is lowered (see for instance \cite{Ishida1998,Miyake2017}). UTe$_2$ follows this law, since $\chi$ is maximum at $T_\chi^{max}=35$~K and a metamagnetic transition occurs at $\mu_0H_m=34$~T for $\mathbf{H}\parallel\mathbf{b}$ \cite{Knafo2019,Miyake2019}. Here, we have further found that the signatures of the CPM boundaries, i.e., a pseudo-gap feature in $\chi$ for $T<T_\chi^{max}$ and the first-order jumps in $M$ and $\Delta L_b/L_b$ at $H_m$, are progressively attenuated for a field angle $\theta<\theta_{QCEP}$, before disappearing beyond a QCEP for $\theta>\theta_{QCEP}$. This QCEP probably results from a subtle balance between the magnetic interactions and the anisotropy of the fluctuating magnetic moments, which were recently shown to have components along the two directions $\mathbf{a}$ and $\mathbf{c}$ \cite{Wang2026}. These experimental findings appeal for a theoretical modeling of the anisotropic effects of a magnetic field on the magnetic fluctuations in UTe$_2$.

A challenge is to determine the relationship between the properties of the metamagnetic transition and the domains of stability of the field-induced superconducting phases SC2 and SC-PPM. Electrical-resistivity \cite{Knafo2019,Knafo2021a,Thebault2022,Thebault2026}, heat-capacity \cite{Rosuel2023a}, magnetization \cite{Miyake2021}, and NMR relaxation-rate \cite{Tokunaga2023} measurements showed an enhancement of the magnetic fluctuations at the metamagnetic field $H_m$ for $\mathbf{H}\parallel\mathbf{b}$. From electrical-resistivity measurements under magnetic fields tilted from $\mathbf{b}$ toward $\mathbf{c}$, magnetic fluctuations peaked at $H_m$ were found to be further enhanced near the onset of SC-PPM for $\theta\simeq30-35^\circ$, indicating their role for the superconducting mechanism \cite{Thebault2026}. At field angles $\theta>35^\circ$, a decrease of the intensity of the critical magnetic fluctuations peaked at $H_m$ suggests that the QCEP reached at $\theta_{QCEP}\simeq70^\circ$ and $\mu_0H_{QCEP}\simeq75$~T may be the place of collapsing critical magnetic fluctuations. In the ferromagnet URhGe, superconductivity is also induced near a metamagnetic transition for $\mathbf{H}\parallel\mathbf{b}$ \cite{Levy2005}, and magnetic fields tilted away from the direction $\mathbf{b}$ lead to the destabilization of superconducting beyond a QCEP \cite{Levy2007}, at which the magnetic fluctuations were found to be reduced \cite{Tokunaga2015}. For UTe$_2$, a demonstration that the QCEP acts as a destabilizing or reinforcing force for field-induced superconductivity is needed. Knowing that, for $\mathbf{H}\parallel\mathbf{b}$, the first-order metamagnetic transition at $H_m$ terminates at the temperature $T_{CEP}\simeq7$~K \cite{Knafo2019,Miyake2019,Miyake2021}, further experiments under wide ranges of temperatures and tilted magnetic fields are now requested to determine if $T_{CEP}$ decreases with increasing field-angle $\theta$ before vanishing at the QCEP. Electrical-transport, thermodynamic and NMR measurements could allow investigating whether the QCEP is the place of enhanced or collapsing quantum critical magnetic fluctuations, with the aim to clarify its relations with SC2 and SC-PPM. In addition, the comparison of the magnetization and lengths variations along the three crystallographic directions $\mathbf{a}$, $\mathbf{b}$, and $\mathbf{c}$ measured in tilted magnetic fields through $H_m$ may help accessing information about the anisotropy of the magnetic properties, via the uniaxial-pressure dependencies of the metamagnetic and QCEP fields extracted using thermodynamic Clapeyron relations (see \cite{Miyake2022}).

\section*{Acknowledgements}

High-magnetic field experiments have been performed at the International MegaGauss Science Laboratory from the Institute for Solid State Physics in Kashiwa. We acknowledge financial support from the French National Research Agency collaborative research projects FRESCO No. ANR-20-CE30-0020, the JSPS KAKENHI (JP22H04933, JP20KK0061, JP24H01641, JP25H01249, JP24H01601, JP23H04862), the ICC-IMR and the GIMRT.


\newpage

\onecolumngrid

\renewcommand\thefigure{S\arabic{figure}}
\renewcommand{\theequation}{S\arabic{equation}}
\renewcommand{\thetable}{S\arabic{table}}
\renewcommand{\thepage}{S\arabic{page}}
\renewcommand{\bibnumfmt}[1]{[S#1]}
\renewcommand{\citenumfont}[1]{S#1}
\setcounter{figure}{0}
\renewcommand{\thesection}{S\arabic{section}}
\renewcommand{\thesubsection}{S\arabic{subsection}}

\vspace{15cm}
\begin{center}
\large {\textbf {Supplemental Material:\\
Thermodynamic signatures of a nearby quantum critical endpoint\\ in UTe$_2$ under intense magnetic field}}
\end{center}
\vspace{1cm}

\setcounter{page}{1}

In this Supplemental Material, we present complementary plots of our magnetization, magnetostriction and magnetic susceptibility data. Focus is given on the variations in the magnetization and relative length at the metamagnetic transition, and on the low-temperature increase in the temperature-derivative of the magnetic susceptibility.

\author{M. Gen}
\affiliation{Institute for Solid State Physics, University of Tokyo, Kashiwa 277-8581, Japan}
\author{K. Somesh}
\affiliation{Laboratoire National des Champs Magn\'{e}tiques Intenses - EMFL, CNRS , Univ. Toulouse, Univ. Grenoble Alpes, INSA-T, Toulouse, France}
\author{A. Miyake}
\affiliation{Institute for Materials Research, Tohoku University, Ibaraki 311-1313, Japan}
\author{D. Li}
\affiliation{Institute for Materials Research, Tohoku University, Ibaraki 311-1313, Japan}
\author{Z. Yang}
\affiliation{Institute for Solid State Physics, University of Tokyo, Kashiwa 277-8581, Japan}
\author{A. Miyata}
\affiliation{Institute for Solid State Physics, University of Tokyo, Kashiwa 277-8581, Japan}
\author{Y. Kohama}
\affiliation{Institute for Solid State Physics, University of Tokyo, Kashiwa 277-8581, Japan}
\author{M. Kondo}
\affiliation{Institute for Solid State Physics, University of Tokyo, Kashiwa 277-8581, Japan}
\author{M. Tokunaga}
\affiliation{Institute for Solid State Physics, University of Tokyo, Kashiwa 277-8581, Japan}
\author{D. Aoki}
\affiliation{Institute for Materials Research, Tohoku University, Ibaraki 311-1313, Japan}
\author{W. Knafo}
\affiliation{Laboratoire National des Champs Magn\'{e}tiques Intenses - EMFL, CNRS , Univ. Toulouse, Univ. Grenoble Alpes, INSA-T, Toulouse, France}
\affiliation{Institute for Materials Research, Tohoku University, Ibaraki 311-1313, Japan}

\newpage
\setcounter{page}{1}

\renewcommand\thefigure{S\arabic{figure}}
\renewcommand{\theequation}{S\arabic{equation}}
\renewcommand{\thetable}{S\arabic{table}}
\renewcommand{\thepage}{S\arabic{page}}
\setcounter{figure}{0}
\onecolumngrid

\begin{figure*}[t]
\includegraphics[width=\textwidth]{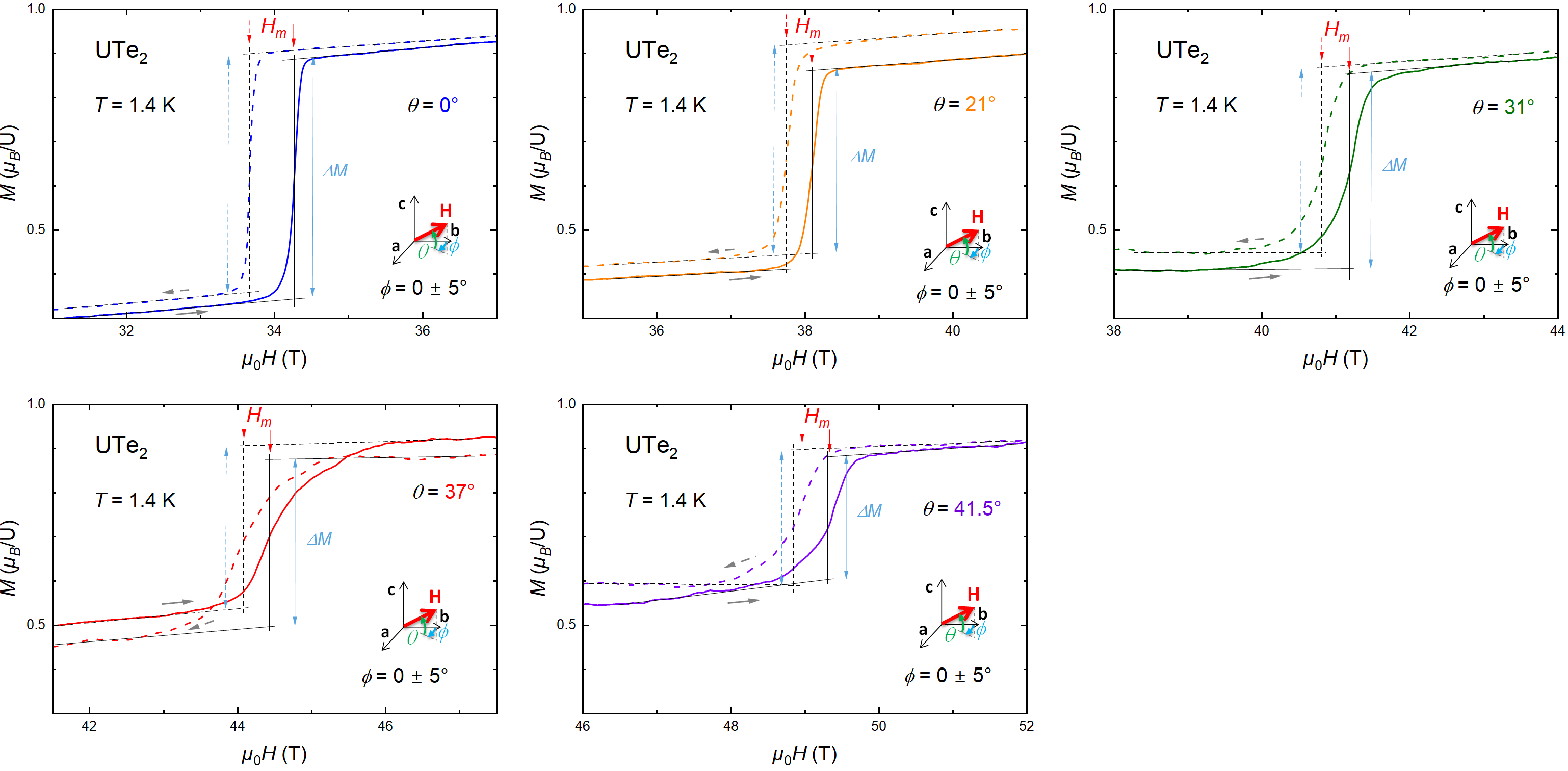}
\caption{\label{FigS1} Magnetic-field variation of the magnetization $M$ at field angles $\theta$ from 0 to $42^\circ$ and temperature $T=1.4$~K, for up (full lines) and down (dashed lines) sweeps of pulsed magnetic fields, and extraction of the metamagnetic field $H_m$  and of the magnetization jump $\Delta M$ at $H_m$.}
\end{figure*}

\begin{figure*}[t]
\includegraphics[width=\textwidth]{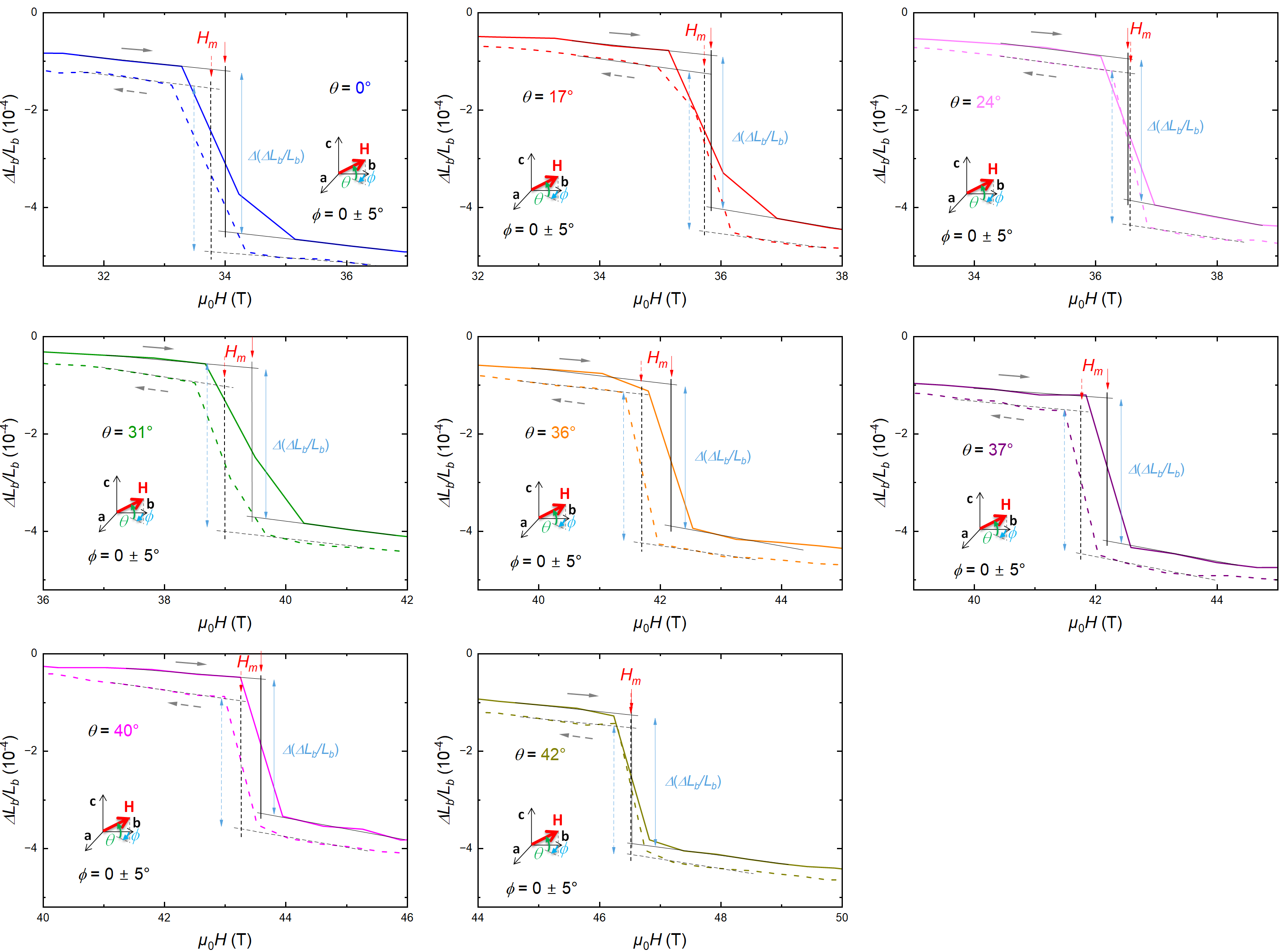}
\caption{\label{FigS2} Magnetic-field variation of the relative length $\Delta L_b/L_b$, where $L_b$ is the sample length along the direction $\mathbf{b}$ and $\Delta L_b=L_b-L_b(H=0)$, at field angles $\theta$ from 0 to $42^\circ$ and temperature $T=1.4$~K, for up (full lines) and down (dashed lines) sweeps of pulsed magnetic fields, and extraction of the metamagnetic field $H_m$  and of the magnetostriction jumps $\Delta (\Delta L_b/L_b)$ at $H_m$.}
\end{figure*}

\begin{figure*}[t]
\includegraphics[width=0.9\textwidth]{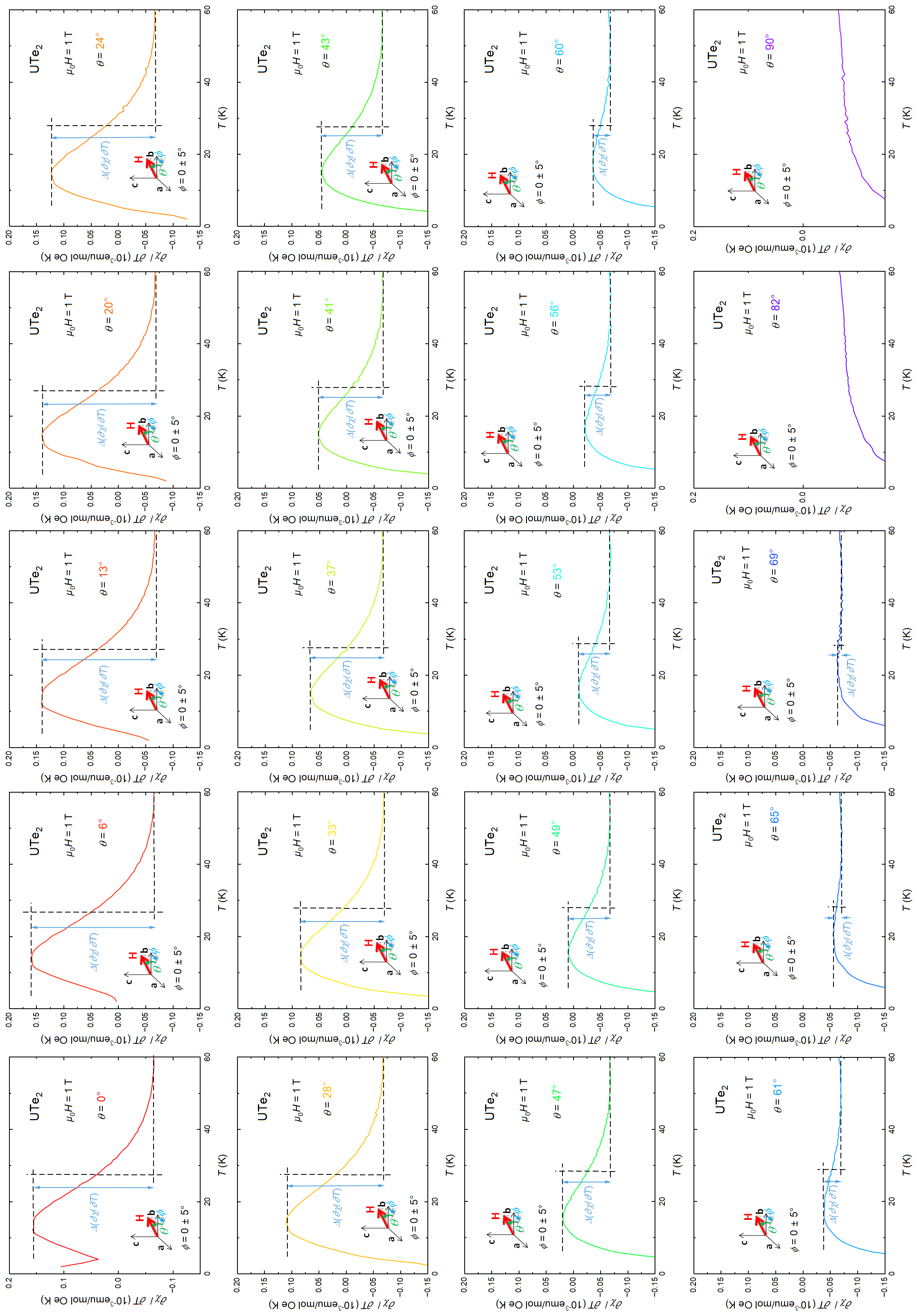}
\caption{\label{FigS3} Temperature variation of the temperature-derivative of the magnetic susceptibility $\partial\chi/\partial T$ measured in magnetic fields of $1$~T tilted by angles $0\leq\theta\leq90^\circ$ from $\textbf{b}$ to $\textbf{c}$, and extraction of the jumps $\Delta (\partial\chi/\partial T)$.}
\end{figure*}

\end{document}